\documentclass[aps,prd,groupedaddress,]{revtex4}
\usepackage{latexsym}
\usepackage[dvips]{graphicx}

\newcommand{\eq}{\begin{equation}}
\newcommand{\feq}{\end{equation}}
\newcommand{\eqn}{\begin{eqnarray}}
\newcommand{\feqn}{\end{eqnarray}}
\newcommand{\arr}{\begin{eqnarray*}}
\newcommand{\farr}{\end{eqnarray*}}
\newcommand{\beq}{\begin{equation}}
\newcommand{\eeq}{\end{equation}}
\newcommand{\bea}{\begin{eqnarray}}
\newcommand{\eea}{\end{eqnarray}}

\def\beq{\begin{equation}}
\def\eeq{\end{equation}}
\def\feq{\end{equation}}
\def\bea{\begin{eqnarray}}
\def\eea{\end{eqnarray}}
\def\bc{\begin{displaymath}}
\def\ec{\end{displaymath}}
\def\lb{\label}

\def\ep{\epsilon}

\def\lb{\label}

\begin{document}
\title{Induced gravity and entanglement entropy of 2D black holes 
\footnote{Talk given at the conference {\it 
From Quantum to Emergent Gravity: Theory and Phenomenology}\\
		June 11-15 2007,
		Trieste, Italy}}

\author{Mariano Cadoni}
\email{mariano.cadoni@ca.infn.it}
\affiliation {\it  Dipartimento di Fisica, Universit\`a di Cagliari \\
and {\it I.N.F.N., Sezione di Cagliari,} \\
{\it Cittadella Universitaria, 09042 Monserrato (Italy)}}

\hfill
\begin{abstract}
    Using the fact that 2D Newton constant is wholly induced by a
conformal field theory,  we derive  a formula for the entanglement entropy 
of the anti-de Sitter black hole 
in two spacetime dimensions.  The  leading term in the large black 
hole mass expansion of our formula reproduces exactly the 
Bekenstein-Hawking  entropy  S$_{BH}$, whereas  the 
subleading term behaves as ln S$_{BH}$. 
This subleading term has the universal form 
typical for the entanglement entropy of physical systems described by 
 effective  conformal fields theories (e.g. one-dimensional 
 statistical models at the critical point). 
\end{abstract}
\hfill

\maketitle

%\FullConference{From Quantum to Emergent Gravity: Theory and Phenomenology\\
	%	June 11-15 2007\\
	%	Trieste, Italy}

\section{Introduction}
 
There have always been  two opposite points of view about the origin 
of Bekenstein-Hawking (BH) black hole entropy. On the one side we 
have the assumption of 
an underlying unitary theory of quantum gravity (QG) and the usual 
statistical paradigm explaining the BH entropy in terms of a 
microstate gas. On the opposite side we do not have 
necessarily  an 
 underlying unitary theory 
and  
BH entropy has to be  thought as  a fundamental feature of QG. 
Entanglement entropy (EE) is a notion which could be  very useful  to
settle down  the controversy. 

Quantum entanglement  is a fundamental feature of quantum systems.  
It is related to the existence of correlations between parts of the system. 
Its investigation has become increasingly relevant in many 
research areas. 
It is an old conjecture, first proposed by 'tHooft,  that black hole entropy 
may be explained in terms of the entanglement  of the quantum state of 
matter fields in the  black hole geometry \cite{'tHooft:1984re,
Bombelli:1986rw,Frolov:1993ym,
Fiola:1994ir,Belgiorno:1995xc,Hawking:2000da,Maldacena:2001kr,
 Brustein:2005vx,Emparan:2006ni,Valtancoli:2006wv}.  
Support to this conjecture comes from the fact that both the EE 
 of matter fields and the BH entropy depend on 
the area of the boundary region. However, any attempt to 
explain the BH entropy as due to quantum entanglement runs into
conceptual and technical difficulties. 

The usual statistical interpretation of the BH entropy 
is conceptually very different from the EE, 
which measures the observer's lack of information about the quantum 
state of the system in a inaccessible region of spacetime.
Moreover, the EE depends both on the number of species $n_{s}$ of the 
matter fields, whose entanglement should reproduce the BH entropy and 
on the value of the UV cutoff $\epsilon$ arising owing to the presence of a sharp 
boundary between the accessible and inaccessible regions of the 
spacetime. Conversely, the BH entropy is meant to be universal,  
independent both from $n_{s}$ and $\epsilon$.
Some conceptual problems look  somehow milder using Sakharov's 
induced gravity approach 
\cite{Jacobson:1994iw,Frolov:1996aj,Frolov:1997up}, but the problem of the 
dependence 
on $n_{s}$ and $\epsilon$ is still there.

Recently, there have been considerable advances in the understanding of the 
EE entropy  in field theory and condensed matter system.
Exact formulas have been derived for the EE entropy of 2D CFTs 
\cite{Holzhey:1994we,Calabrese:2004eu}.
It has been shown that  EE plays a 
important role in condensed matter systems, where it helps to understand quantum 
phases of matter (e.g spin chains and quantum 
liquids)\cite{Vidal:2002rm,its,Kitaev:2005dm,Latorre:2004pk,korepin}.
Entanglement  entropy turned out to be  an useful concept also for 
 investigating general features of  quantum field theory (QFT), in particular  
two-dimensional
conformal field theory (CFT),  the Anti-de Sitter/conformal 
field theory (AdS/CFT) correspondence and the existence of c-theorems 
\cite{Casini:2004bw,Fursaev:2006ng,
Solodukhin:2006xv,Ryu:2006bv,Ryu:2006ef,Hubeny:2007xt}.

One is therefore tempted to use this new ideas and  improved 
understanding of the EE to tackle the old problems related to the EE 
entropy of black holes.
In this contribution, which is mainly based on Ref. 
\cite{Cadoni:2007vf}, we will show that this is possible at least in the case of 
two-dimensional AdS 
black holes.  We will 
derive an expression for the black hole EE that in the large black 
hole mass limit   reproduces exactly the 
BH entropy. Moreover, we will show that the subleading term has the 
universal behavior typical for CFTs and in particular for critical phenomena. 
The reason of this success is related to the 
peculiarities  of 2D AdS gravity, namely the existence of an 
AdS/CFT correspondence and the fact that  2D Newton constant can be 
considered as 
wholly induced by  quantum fluctuations of the dual CFT.

\section{Entanglement entropy for 2D CFTs}

Most of the progress in understanding the EE in QFT has been 
achieved in the case of 2D CFT. 
If only a spacelike slice $Q$ (of length $\Sigma$) of our 2D universe, 
endowed with an IR cutoff $\Lambda$, is accessible for measurement we have to 
trace over the degrees of freedom in the complementary region $P$
(of length $\Lambda -\Sigma$).
The entanglement entropy   originated 
by tracing over the unobservable degrees of freedom is given by the  
Von Neumann entropy 
$S_{ent}=- 
Tr_{Q}{\hat \rho}_{Q}\ln\hat\rho_{Q}$, 
where 
the reduced density matrix $\hat\rho_{Q}=Tr_{P}\hat\rho$ is obtained by tracing 
the density  matrix $\hat \rho$ over  states in the region $P$.
The resulting  EE for the ground state of the 2D CFT
has been first calculated 
by  Holzhey, Larsen and Wilckzek \cite{Holzhey:1994we}.
Owing to the contribution of 
localized excitations 
arbitrarily near to the boundary the entanglement entropy diverges. 
Introducing an ultraviolet cutoff $\epsilon$,  the regularized 
entanglement entropy turns out to be 
\beq\lb{f5}
S_{ent}= \frac{c+\bar c}{6}\ln\left(\frac{\Lambda}{\epsilon 
\pi}\sin\frac{\pi \Sigma}{\Lambda}\right),
\feq
where $c$ and $\bar c$ are the central charges of the 2D CFT.

Characterizing features of the expression (\ref{f5}) for the EE are:\\
$a)$ Subadditivity\\
$b)$ Invariance under the 
transformation which 
exchanges  the inside and outside regions
\beq\lb{f6}
\Sigma\to \Lambda-\Sigma.
\feq
$c)$  $S_{ent}$ is not a monotonic function of $\Sigma$, but 
increases and reaches its maximum for $\Sigma=\Lambda/2$ and then 
decreases as $\Sigma$ increases further. 

\section{2D AdS black hole and the AdS/CFT correspondence}

 2D AdS black holes
are   classical solutions of a 2D (dilaton) gravity theory with action
$A=\int d^{2}x \sqrt{g}\Phi(R+2/L^{2})$,
where the length $L$ is related to   
cosmological constant of the AdS spacetime ($\lambda=1/L^{2}$) and 
$\Phi$ is the dilaton.
In the Schwarzschild gauge the 2D AdS  black hole solutions 
are \cite{Cadoni:1994uf},
\beq\lb{f2}
ds^{2}= -\frac{1}{L^{2}}\left(r^{2}-r_{h}¥^{2}¥\right)dt^{2}+
L^{2}\left({r^{2}}- r_{h}^{2}¥\right)^{-1}¥dr^{2},\quad \Phi= 
\frac{r}{GL},
\feq
where $G$ is the dimensionless 2D Newton constant and  $r_{h}¥$ is 
the horizon  radius.

The black hole mass, temperature and  Bekenstein-Hawking entropy are 
given by  \cite{Cadoni:1994uf} 
\beq\lb{f3a}
M= \frac{r_{h}¥^{2}}{2GL^{3}},\quad T=\frac{r_{h}}{2\pi L^{2}},\quad
S_{BH}¥=\frac{ 2\pi r_{h}}{GL} .
\feq
Setting  $r_{h}=0$ in Eq. (\ref{f2})  we have the  AdS black hole ground 
state  with zero mass, temperature 
and entropy. The AdS black hole (\ref{f2}) can be considered as the 
thermalization of the  ground state  solution \cite{Cadoni:1994uf}. 
 
The 2D black hole has a dual description in 
terms of a
 CFT with central charge 
 \cite{Cadoni:1998sg,Cadoni:1999ja,Cadoni:2000kr,Cadoni:2000fq}
\beq\lb{f4}
c= \frac{12}{G}.
\feq
The dual CFT can have  both the form of a 2D \cite{Cadoni:2000kr,Cadoni:2000fq}
or a 1D
\cite{Cadoni:1998sg,Cadoni:1999ja} conformal field  theory.
This  AdS$_{2}$/CFT$_{2}$ ( or AdS$_{2}$/CFT$_{1}$)
correspondence has been used to 
give a microscopical meaning to the thermodynamical 
entropy of 2D AdS black holes. The BH entropy (\ref{f3a})  has been reproduced 
by counting states in the  dual CFT.

\section{Entanglement entropy of the 2D AdS black hole}

It has been  observed that in two dimensions  black hole entropy 
can be ascribed to quantum entanglement if 2D Newton 
constant  is wholly induced by  quantum fluctuations of matter 
\cite{Fiola:1994ir,Susskind:1994sm,
Frolov:1996aj,Frolov:1997up}.
But Eq. (\ref{f4}) (or more in general the 
AdS$_{2}$/CFT$_{2}$ correspondence) tells us that the 2D Newton constant 
is 
induced by quantum fluctuations of the dual CFT. 
It follows that in the semiclassical approximation 
the black hole  EE  should be identified with  
the entanglement  entropy of the vacuum 
for the 2D CFT 
of central charge given by Eq. (\ref{f4}) in the  black 
hole geometry (\ref{f2}). 
The black hole exterior and interior should be identified 
respectively with the observable 
region $Q$ (of length  $\Sigma$) and 
unobservable region $P$ (length $\Lambda -\Sigma$) of the 2D 
universe described above, where the 
CFT degrees of freedom live.

However, two obstacles prevent a direct application  
of Eq. (\ref{f5}):\\
$1)$ Eq.  (\ref{f5}) holds for a 2D flat 
spacetime, whereas we  are dealing with  a curved 2D background.\\
$2)$  The calculations leading to Eq. (\ref{f5}) are performed for 
a spacelike slice $Q$, whereas 
in the  black hole case, 
owing to the coordinate singularities  at  $r=r_{h}$ (the horizon) and $r=\infty$
(the timelike asymptotic boundary of the AdS spacetime),
there is  no global 
notion of a spacelike coordinate. \\
Owing to these geometrical features, in the black hole
case we   cannot give a direct meaning to {\sl both} the measures 
$\Sigma$ and $(\Lambda-\Sigma)$ of the subsystems  $Q,P$
and  invariance under the transformation 
(\ref{f6}) becomes  meaningless.

The second difficulty can be circumvented using  appropriate 
coordinate system and  regularization procedure,  the first  using 
instead of Eq. (\ref{f5}) the formula derived by Fiola et al. 
\cite{Fiola:1994ir}, which gives the EE  of the vacuum of matter 
fields in the case of 
a curved  background\footnote{Notice that we are using the 
formula of Ref. \cite{Fiola:1994ir} with  reversed sign.
The AdS black hole has to be considered as the  vacuum seen 
by the observer using the black hole coordinates.
This observer sees the the CFT vacuum as filled with 
thermal radiation with {\sl negative} flux \cite{Cadoni:1994uf}.}, 
\beq\lb{f7}
S_{ent}=  -\frac{c}{6}\left( \rho|_{boundary}¥- 
\ln\frac{\epsilon}{\Lambda }\right),
\feq
where
 $c$ is the 
central charge (\ref{f4}) and
$\exp(2\rho)$ is the conformal factor of the metric in the 
coordinate system used to define the CFT vacuum.
In this coordinate system the black hole metric (\ref{f2}) takes the 
form \cite{Cadoni:1994uf}
\beq\lb{f9}
ds^{2}= 
\frac{r_{h}¥^{2}}{L^{2}}\frac{\left(-dt^{2}+d\sigma^{2}\right)}
{\sinh^{2}(\frac{r_{h}¥\sigma}{L^{2}¥})}.
\feq
Notice that in Eq. (\ref{f7}) we have contributions from only one sector
(e.g. right movers) of the 
CFT. In Ref. \cite{Cadoni:2000kr,Cadoni:2000fq} it has been shown that the 2D  AdS black hole 
is dual to an open string with appropriate boundary conditions.
These boundary conditions are such that only one sector of the 
CFT$_{2}$ is present. 

\begin{figure}
  \includegraphics[width=300pt]{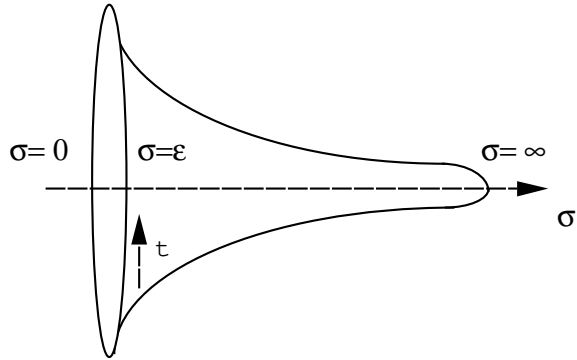}\\
  \caption{ Regularized euclidean instanton corresponding to the 2D 
  AdS black hole in the coordinate system $(t,\sigma)$ covering only 
  the black hole exterior. The euclidean time is periodic.
  The point $\sigma=\infty$ correspond to the 
  black hole horizon. $\sigma=0$ corresponds to the asymptotic 
  timelike boundary of AdS$_{2}$. }\label{instanton}
\end{figure}

The coordinate system $(t,\sigma)$  covers only the black 
hole  exterior. Working in euclidean space the 2D manifold has only a 
boundary at $\sigma=0$, corresponding to $r=\infty$, the timelike 
conformal boundary of the 2D AdS space. 
The conformal factor of the metric (\ref{f9}), hence  the entanglement 
entropy (\ref{f7}), blows up on the 
$\sigma=0$ boundary of the AdS spacetime.    The simplest 
regularization procedure  is to consider a 
regularized boundary at $\sigma=\ep$.
Notice that $\ep$ 
plays the role of a UV cutoff for the coordinate $\sigma$, which is 
the natural spacelike coordinate of the dual CFT. $\ep$  is an IR 
cutoff for the coordinate $r$, which is the natural spacelike 
coordinate for the AdS$_{2}$ black hole.
The regularized euclidean instanton corresponding to the black hole 
(\ref{f9}) is shown in figure  (\ref{instanton}).

The regularized boundary is at finite proper distance from 
the horizon. $\ep$ acts also as IR regulator, making the 
presence of the IR cutoff $\Lambda$ in Eq. (\ref{f7}) redundant.
The EE still depends on the UV cutoff $\ep$.
The AdS/CFT correspondence enable us to identify  $\ep$
as the UV cutoff of the CFT: $\ep\propto L$.
The proportionality factor can be determined by requiring that the 
analytical continuation of Eq. (\ref{f11b}) below  is invariant under the 
transformation (\ref{f6})
(see later). This requirement fixes $\ep=\pi L$.

Putting all together we obtain  from Eq. (\ref{f7}) the EE entropy of 
the 2D CFT in the classical black hole geometry,
\beq\lb{f11b}
S_{ent}^{(bh)}=\frac{c}{6} \ln \left( \frac{L¥}{\pi r_{h}} \sinh 
\frac{\pi 
r_{h}}{L}\right), 
\feq
which has to be identified with the EE of the 2D AdS black hole.
The   entanglement entropy (\ref{f11b}) has the expected  behavior as a 
function of the horizon radius $r_{h}$ or, equivalently, of the black 
hole mass $M$. $S_{ent}^{(bh)}$ becomes  zero in the AdS ground 
state, $r_{h}¥=0$ ($M=0$), whereas it grows monotonically for $r_{h}>0$ 
($M>0$).

Let us now 
consider the large mass  behavior $r_{h}>>L$  (macroscopic black 
holes) of Eq. (\ref{f11b}):

\beq\lb{f12}
S_{ent}^{(bh)}=  \frac{2\pi r_{h}}{GL}-  \frac{2}{G}\ln\frac{r_{h}}{L}
+O(1)= S_{BH}- \frac{2}{G}\ln S_{BH} +O(1).
\feq
We have obtained the remarkable result that the leading term  in the 
large mass expansion of the black hole entanglement entropy reproduces exactly 
the Bekenstein-Hawking  entropy. In this regime the thermal length 
$\beta=1/T<<L$, thermal correlations dominate and the EE becomes 
Gibbs entropy. 
The subleading term  has the universally predicted  $\ln S_{BH}$ 
behavior for the quantum corrections to the Bekenstein-Hawking result.
Notice that, although it is an universally accepted result 
that the quantum corrections to 
the BH entropy behave as $\ln S_{BH}$ 
\cite{Fursaev:1994te,Mann:1997hm,Kaul:2000kf,Carlip:2000nv,Ghosh:1994wb,
Mukherji:2002de,Setare:2003vv,Domagala:2004jt,Medved:2004eh,Grumiller:2005vy},
there is no 
general agreement about the value of the prefactor of this term. 
Equation (\ref{f12}) fixes the prefactor of  $\ln S_{BH}$ in terms 
of the  2D Newton constant. This result 
contradicts some previous results  supporting a 
$G$-independent  value of the prefactor.
Our result is consistent with the  approach followed in this paper,
which considers 2D gravity as induced from the quantum fluctuations 
of a  CFT with central charge $12/G$. 
The subleading term in Eq. (\ref{f12}) has also the
universal 
behavior shared  with other  systems described by 2D QFTs,
such as one-dimensional 
statistical models near  to the critical point (with  the black hole 
radius $r_{h}$
corresponding to the  correlation length) or free  scalars 
fields  
\cite{Calabrese:2004eu,Fursaev:2006ng}.

For $\beta>>L$ quantum correlations dominate but we expect our 
semiclassical description, which also neglects back-reaction,
to break down  at $\beta\sim L$. In this 
regime it is likely that a phase transition, analogue to the 
Hawking-Page transition in 4D takes place.

The black hole EE (\ref{f11b}) differs from the  CFT  formula (\ref{f5})
by the exchange of  hyperbolic with 
trigonometric  functions. 
The appearance of hyperbolic functions  solve the  problems concerning 
the application of  formula (\ref {f5}) to the black hole case.
It allows for monotonic increasing of $S_{ent}^{(bh)}(r_{h})¥$, 
eliminating the  unphysical decreasing  
behavior of $S_{ent}(\Sigma)$ in the region $\Sigma> \Lambda/2$.

\begin{figure}
  \includegraphics[width=300pt]{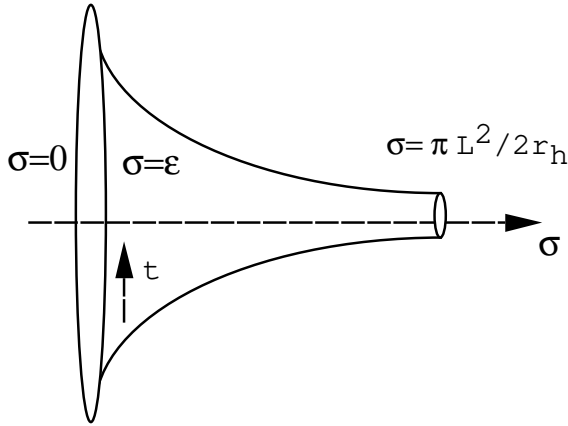}\\
  \caption{ Regularized euclidean instanton corresponding to the 2D 
  AdS black hole with negative mass. The euclidean time is periodic.
  The point $\sigma=\pi L^{2}¥/2r_{h}¥$ 
  corresponds to the 
  black hole singularity at $r=0$. $\sigma=0$ corresponds to the asymptotic 
  timelike boundary of AdS$_{2}$. }\label{instanton1}
\end{figure}

On the other hand the presence of hyperbolic instead of trigonometric 
function indicates that Eq. (\ref{f5}) can be obtained as the 
analytic continuation $r_{h}\to i r_{h}$ of our formula (\ref{f11b}), 
i.e  by considering an AdS black hole with negative mass.
The analytically continued black hole solution is given by Eq. 
(\ref{f2}) with $r_{h}¥^{2}<0$. In the conformal gauge the solution 
reads now $ds^{2}= [r_{h}¥/(L\sin(r_{h}¥\sigma /L^{2}¥)]^{2}¥(-dt^{2}+d\sigma^{2})$.
The corresponding regularized euclidean 
instanton  is shown in Fig. (\ref{instanton1}). 
In terms of the 2D CFT we have to trace over 
the degrees of freedom outside the spacelike slice $\ep <\sigma< 
\pi L^{2}¥/2r_{h}¥$.  Applying Eq. (\ref{f7} to the case of a spacelike slice  with 
two boundary points and 
 redefining appropriately the UV cutoff $\epsilon$, we get
\beq\lb{f15}
S_{ent}= \frac{c}{6} \ln \left(\frac{\Lambda}{\pi \epsilon} \sin\frac{\pi 
r_{h}}{\Lambda}\right).
\feq
Thus, the entanglement entropy of the 2D CFT in the curved background 
given by the AdS black hole of negative mass has exactly the form 
given by Eq. (\ref{f5}) with the horizon radius $r_{h}$ playing the 
role of $\Sigma$.
The requirement that  equation (\ref{f15}) is the analytic 
 continuation of Eq. (\ref{f11b})  fixes, as previously 
 anticipated, the proportionality factor between $\ep$ and $L$ in the 
 calculations leading to Eq. (\ref{f11b}).

\section{Conclusions}
Thanks to the AdS/CFT correspondence we were able to 
solve the number of species and UV cutoff problem and to 
derive a formula for the entanglement 
 entropy of 2D  AdS black holes that has nice striking features.
 The leading term in the large black hole mass expansion reproduces 
 exactly the BH entropy. The subleading term has the right $\ln 
 S_{BH}$, behavior of the quantum corrections to the BH 
 formula and represents an universal term typical of CFTs.
 
The picture that emerges is intriguing but to a large extend 
expected: the black hole EE entropy reduces to the 
Bekenstein-Hawking entropy only for macroscopic black holes when 
thermal correlations dominate. Away from this regime the EE codifies 
information about the full quantum gravity description.
Black hole EE could be therefore used to gain information about  quantum 
gravity phase transition (e.g Hawking-Page-like phase transition).
In particular,  it could be used to describe
the emergence of a spacetime description from the 
microscopic  theory, in the same way as it is used to 
gain information about quantum phase transitions in condensed matter 
systems.
Obviously, in the full quantum gravity  regime $\beta>>L$ our 
description breaks down and  the full
underlying microscopic theory ( e.g string theory on AdS$_{2}$ times 
some compact manifold) has to be used.

Generalization of our approach to higher dimensional gravity theory 
would be more then welcomed. 
Our results rely heavily on  peculiarities of 2D AdS gravity, namely 
the  existence of an AdS/CFT correspondence and on the fact that 2D 
Newton constant arises from quantum fluctuation 
of the dual CFT.
The generalization of our approach to higher dimensional gravity 
theories is therefore far from being trivial.

\end{document}